\documentclass[letter]{ptptex}

\usepackage{graphicx}

\title{
Gravitational microlensing in modified gravity theories: 
Inverse-square theorem 
}

\author{
Hideki \textsc{Asada}%
}

\inst{
Faculty of Science and Technology, Hirosaki University, 
Hirosaki 036-8561, Japan
}

\abst{
Microlensing studies are usually based on the lens equation that 
is valid only to the first order in the gravitational constant $G$ 
and lens mass $M$. 
We consider corrections to the conventional lens equation 
in terms of differentiable functions, so that they can express 
not only the second-order effects of $GM$ 
in general relativity but also 
modified gravity theories. 
As a generalization of Ebina et al. 
(Prog. Theor. Phys. {\bf 104} (2000) 1317), 
we show that, 
provided that the spacetime is static, spherically symmetric 
and asymptotically flat, 
the total amplification by microlensing 
remains unchanged at the linear order of the correction 
to the deflection angle, 
if and only if the correction takes 
a particular form as the inverse square of the impact parameter, 
whereas the magnification factor for each image is corrected. 
It is concluded that 
the light curve shape by microlensing is inevitably 
changed and will thus allow us to probe modified gravity,  
unless a modification to the deflection angle 
takes the particular form. 
No systematic deviation in microlensing observations 
has been reported. 
For instance, therefore, the Yukawa-type correction 
is constrained as the characteristic length $> 10^{14}$ m. 
}

\begin{document}

\maketitle

\section{Introduction}
Since the pioneering work by Paczynski \cite{Paczynski} 
and the subsequent detections of MACHOs \cite{Alcock,Aubourg},  
microlensing has been one of the most vital areas 
in astrophysics. Now, it plays an important role 
also in searching extra-solar planets \cite{Beaulieu,Gaudi}. 
Light rays are influenced by a local curvature of a spacetime \cite{SEF}. 
It seems natural that the bending angle of light rays 
in modified gravity theories is different from 
that in general relativity. 
Does such a modified bending angle really makes a change 
in microlensing? 
We shall examine this problem in this paper. 

Conventional studies of microlensing are usually 
based on the lens equation that maps the source direction 
into the position on the lens plane, namely the direction of 
each image due to the lens effect \cite{SEF}. 
In particular, the deflection angle is written as 
$4GM/bc^2$, where $G$ is the gravitational constant, 
$M$ is the lens mass, $b$ is the impact parameter of the light ray, 
and $c$ is the speed of light. For the derivation 
of the deflection angle and consequently the lens equation, 
the post-Minkowskian approximation $O(G)$ or the linear-order 
approximation $O(M)$ of the Schwarzschild metric is employed. 

Do second-order relativistic corrections 
affect microlensing through the amplification factor? 
This has been already answered \cite{EOAK}. 
The total amplification remains unchanged at $O(G^2 M^2)$, 
whereas the magnification factor for each image is corrected 
at this order. 

The main purpose of this paper is to generalize 
their relativistic result. 
In particular, theoretical models beyond the theory 
of general relativity have attracted a lot of interests 
last decades, mostly motivated by the dark energy and 
dark matter problems. 
In this letter, therefore, we shall reexamine 
the amplification of lensed images by taking account of 
corrections in a rather general form, 
where we assume static, spherically symmetric 
spacetimes. 

We use the units of $c=1$ but keep $G$ 
in order to make iterative calculations clearer. 

\section{Corrections to the lens equation and magnification}
\subsection{Standard form of lens equation and magnification} 
Let us begin by briefly summarizing the derivation 
of the total amplification for microlensing.\cite{Paczynski} 
We denote distances from the observer to the lens, from the observer 
to the source, and from the lens to the source 
as $D_L$, $D_S$ and $D_{LS}$, respectively. 
The source and image position angles are denoted 
by $\theta_S$ and $\theta_I$, respectively. 
The lens equation is written as 
\begin{equation}
\theta_S=\theta_I-\frac{D_{LS}}{D_S}\alpha , 
\end{equation}
where we used the thin-lens approximation. 
This gives us a mapping between the lens and source planes. 
We can choose $\theta_S \geq 0$ without loss of generality. 
For a point mass lens 
(generally a spherically symmetric lens), 
the deflection angle $\alpha$ at $O(G)$ becomes   
\begin{equation}
\alpha=\frac{4GM}{b}. 
\end{equation}
Hence, the lens equation is
\begin{equation}
 \theta_S=\theta_I-\frac{\theta_E^2}{\theta_I}, 
\end{equation}
where $\theta_E$ is the angular radius of the  Einstein ring as 
\begin{equation}
\theta_E=\sqrt{\frac{4GMD_{LS}}{D_L D_S}}.   
\label{eq:Einsteinring}
\end{equation} 
For microlensing in our galaxy, the radius is typically 
\begin{equation}
\theta_E \sim 10^{-3} \Bigl( \frac{M}{M_{\odot}} \Bigr)^{1/2} 
\Bigl( \frac{10\mbox{kpc}}{D_{S}} \Bigr)^{1/2} \mbox{arcsec.} , 
\end{equation}
where we assumed $D_L \sim D_{LS}$. 

The lens equation is solved easily as 
\begin{equation}
\theta_{\pm}=\frac12(\theta_S \pm \sqrt{\theta_S^2+4 \theta_E^2}) . 
\end{equation}
We obtain the amplification due to gravitational lensing as  
\begin{equation}
A=\Bigl|\frac{\theta_I}{\theta_S}\frac{d\theta_I}{d\theta_S}\Bigr| 
 =  \frac{1}{\left|1 - \left(\frac{\theta_E}{\theta_I}\right)^4\right|}, 
\label{eq:amp}
\end{equation}
which is a function of the image position $\theta_I$. 

For each image at $\theta_{\pm}$, the amplification factor becomes 
\begin{equation}
A_{\pm}=\frac{u^2+2}{2u\sqrt{u^2+4}} \pm \frac12 , 
\end{equation}
where $u$ denotes $\theta_S/\theta_E$, the source position 
in units of the Einstein ring radius. 
For microlensing events, the angular separation between the images 
is too small to measure. 
All we can measure is the total amplification that 
is expressed as  
\begin{equation}
A_{\mbox{total}}=A_{+} + A_{-}=\frac{u^2+2}{u\sqrt{u^2+4}}. 
\label{eq:amplification}
\end{equation}

\subsection{Modified bending angle and 
its effects on magnification} 
Let us take account of some modification 
in the bending angle of light rays. 
We focus on modified gravity theories that consider 
spacetimes to be {\it differentiable} manifolds. 
This implies that the modified bending angle can be expressed 
in terms of differentiable functions. 
We assume also that the lens object produces a static  
and spherically symmetric spacetime, 
for which the bending angle may take a general form as 
\begin{equation}
\alpha = \frac{4GM}{b} \left[ 1 + F(b) \right] , 
\label{alpha2}
\end{equation}
with a differentiable function $F(b)$ denoting modifications. 
This point is contrast to previous works \cite{EOAK,KP2005,KP2006}, 
in which particular forms such as the second-order of mass 
in the Schwarzschild spacetime and the PPN formalism are assumed. 

For specific models, $F(b)$ may be approximated  
in terms of power functions $b^p$ \cite{Asada2008}. 
For instance,  
$p = 1/2$ for the DGP model as one of brane scenarios \cite{DGP}, 
and 
$p = 3/2$ for a candidate of massive graviton theories \cite{MG1,MG2}.  

The lens equation to be solved becomes 
\begin{equation}
\tilde\theta_S=\tilde\theta_I - \tilde\alpha(\tilde\theta_I) , 
\label{eq:lenseq}
\end{equation}
with the modified bending angle as 
\begin{equation}
\tilde\alpha(\tilde\theta_I) \equiv 
\frac{1}{\tilde\theta_I} 
\left( 1 + \varepsilon f(|\tilde\theta_I|) \right) , 
\label{tildealpha}
\end{equation}
where we introduce a nondimensional small quantity $\varepsilon$ 
as the expansion parameter and 
$f(|\tilde\theta_I|)$ 
is an arbitrary differentiable function corresponding to $F(b)$. 
Here, $\tilde\theta_S$ and $\tilde\theta_I$ are normalized 
by $\theta_E$ as $\tilde\theta_S \equiv \theta_S / \theta_E = u$ 
and $\tilde\theta_I \equiv \theta_I / \theta_E$, respectively. 
We should note that 
$f(\tilde\theta_I = +\infty)$ can be absorbed into the mass 
that is defined at the spatial infinity 
and hence $f(\tilde\theta_I = +\infty)$ can be taken as zero. 
Henceforth, we assume $f(\tilde\theta_I) \to 0$ 
as $\tilde\theta_I \to \infty$. 

This lens equation is rewritten as 
\begin{equation}
\tilde\theta_I^2 - \tilde\theta_S \tilde\theta_I 
- 1 - \varepsilon f(|\tilde\theta_I|)
= 0 . 
\label{eq:lenseq2}
\end{equation}

The lens equation $(\ref{eq:lenseq2})$ can be solved iteratively 
in terms of $\varepsilon$. 
For $\tilde\theta_I \geq 0$, the lens equation is rewritten as 
\begin{equation}
\tilde\theta_I^2 - \tilde\theta_S \tilde\theta_I 
- 1 - \varepsilon f(\tilde\theta_I)
= 0 . 
\label{modifiedlenseq+}
\end{equation} 
A perturbative form as 
\begin{equation}
\tilde\theta_{+}^{'} 
= \tilde\theta_{+} + 
\varepsilon\phi_{+} + O(\varepsilon^2) , 
\end{equation} 
is substituted into the image position $\tilde\theta_I$ 
in Eq. $(\ref{modifiedlenseq+})$,  
where the prime denotes a quantity including 
effects by the modified bending angle. 
Here, one can assume $\tilde\theta_{+}^{'} > 0$ 
because of $\tilde\theta_{+} > 0$, 
within the limit that perturbative calculations hold. 
Then, we find the solution as 
\begin{equation}
\tilde\theta_{+}^{'} = \tilde\theta_{+} + 
\varepsilon \frac{f(\tilde\theta_{+})}{\sqrt{\tilde\theta_S^2 +4}} 
+ O(\varepsilon^2) . 
\label{theta+}
\end{equation}

For $\tilde\theta_I < 0$, the lens equation becomes 
\begin{equation}
\tilde\theta_I^2 - \tilde\theta_S \tilde\theta_I 
- 1 - \varepsilon f(-\tilde\theta_I)
= 0 . 
\label{modifiedlenseq-}
\end{equation} 
In the similar manner to the positive case, 
one finds the solution as 
\begin{equation} 
\tilde\theta_{-}^{'} = \tilde\theta_{-} - 
\varepsilon \frac{f(-\tilde\theta_{-})}{\sqrt{\tilde\theta_S^2 +4}} 
+ O(\varepsilon^2) . 
\label{theta-}
\end{equation}

By using Eqs. ($\ref{theta+}$) and ($\ref{theta-}$), 
we obtain the amplification for each image as 
\begin{equation}
A_{\pm}^{'} = \frac{u^2+2}{2u\sqrt{u^2+4}} 
\pm \left( \frac12 + \varepsilon \ell_{\pm} \right) 
+ O(\varepsilon^2) , 
\label{modifiedApm}
\end{equation}
where we define the correction term $\ell_{\pm}$ as 
\begin{equation}
\ell_{\pm} 
\equiv 
\pm 
\frac{\tilde\theta_{\pm}}{\tilde\theta_S}
\frac{d \tilde\theta_{\pm}}{d \tilde\theta_S}
\frac{1}{\sqrt{u^2 +4}}
\left(
\frac{f(\pm \tilde\theta_{\pm})}{\tilde\theta_{\pm}} 
+ \frac{d f(\pm \tilde\theta_{\pm})}{d \tilde\theta_{\pm}} 
\right)
\mp 
f(\pm \tilde\theta_{\pm}) 
\frac{\tilde\theta_S}{(u^2 +4)^{3/2}} . 
\label{ell}
\end{equation}
Here, we assume that the sign of the Jacobian for the lens mapping 
does not change at the linear order of the modified bending angle 
as far as perturbative treatments are valid.
Clearly the modified bending angle leads to 
a change in the amplification 
for each image. 

By using Eq. ($\ref{modifiedApm}$), we obtain 
the total amplification as 
\begin{equation}
A_{\mbox{total}}^{'} = \frac{u^2+2}{u\sqrt{u^2+4}} 
+ \varepsilon (\ell_{+} - \ell_{-}) 
+ O(\varepsilon^2) .  
\label{modifiedAtotal}
\end{equation}

According to Eq. ($\ref{modifiedAtotal}$), 
$\ell_{+} = \ell_{-}$ is equivalent to 
the condition that the total amplification 
for the modified bending angle remains unchanged 
at the linear order. 
By using Eq. ($\ref{ell}$), this condition is rewritten as 
\begin{equation}
\frac{d}{du} 
\frac{\tilde\theta_{+} f(\tilde\theta_{+})}{\sqrt{u^2+4}} 
= 
\frac{d}{du} 
\frac{(-\tilde\theta_{-}) f(-\tilde\theta_{-})}{\sqrt{u^2+4}} , 
\end{equation}
where we should note that $\tilde\theta_{\pm}$ is a function of $u$. 
This is integrated as 
\begin{equation}
\frac{\tilde\theta_{+} f(\tilde\theta_{+}) 
+ \tilde\theta_{-} f(-\tilde\theta_{-})}{\sqrt{u^2+4}}
= const. 
\label{integ}
\end{equation}

For the asymptotically flat case, 
this constant must vanish as shown below. 
In the limit as $u \to \infty$, we have 
$\tilde\theta_{+} \to u \to \infty$ and $\tilde\theta_{-} \to 0$. 
Hence the L.H.S. of Eq. ($\ref{integ}$) becomes
$f(\tilde\theta_{+} = \infty) = 0$ as $u \to \infty$, 
which means that the above constant vanishes. 
Therefore, the condition of $\ell_{+} = \ell_{-}$ is 
expressed as 
\begin{equation}
\tilde\theta_{+} f(\tilde\theta_{+}) 
= 
- \tilde\theta_{-} f(-\tilde\theta_{-}) . 
\end{equation}
By using the identity as 
$\tilde\theta_{+} \tilde\theta_{-} = -1$, 
this is rewritten simply as 
\begin{equation}
g(x) = g\left(\frac{1}{x}\right) , 
\label{condition}
\end{equation}
where we define $g(x) \equiv x f(x)$. 
Let us prove that $g(x)$ is a constant. 
We expand the differentiable function $g(x)$ 
in the Laurent series as 
\begin{equation}
g(x) = \sum_{r=-\infty}^{\infty} a_r x^r, 
\label{g-expansion}
\end{equation}
where $a_r$ is some constant and singular points may exist. 
This gives also $g(1/x)$ as 
\begin{equation}
g\left(\frac{1}{x}\right) = \sum_{r=-\infty}^{\infty} a_{-r} x^r. 
\end{equation}
For these expansions, Eq. ($\ref{condition}$) tells us 
\begin{equation}
a_r = a_{-r} ,  
\end{equation}
which allows us to rearrange the expansion of $g(x)$ as 
\begin{equation}
g(x) = a_0 + a_1 \left( x+\frac{1}{x} \right) 
+ a_2 \left( x^2+\frac{1}{x^2} \right) 
+ \cdots . 
\end{equation}

The asymptotic flatness requires that 
the bending angle vanishes at the spatial infinity. 
Namely, 
$x^{-1} f(x)$ vanishes as $x \to \infty$, 
which means $a_r = 0$ 
for $r \geq 2$.  
Then, $g(x) = a_0 + a_1(x + x^{-1})$ gives the bending angle 
defined by Eq. (\ref{tildealpha}) as 
\begin{equation}
\tilde\alpha(\tilde{b}) 
= \frac{1}{\tilde{b}} 
\left[ 1 + \varepsilon 
\left\{
\frac{a_0}{\tilde{b}} 
+ a_1 \left( 1 + \frac{1}{\tilde{b}^2} \right)
\right\} 
\right] . 
\end{equation}
However, the two terms with $a_1$ are rewritten as 
\begin{equation}
1 + \frac{\theta_E^2}{\theta_I^2} 
= 1 + \frac{4GMD_LD_{LS}}{D_Sb^2} . 
\label{a1}
\end{equation}
This expression makes no sense, 
because the bending angle is calculated by assuming 
the observer and source located at the null infinity 
and hence it includes neither $D_{LS}$, $D_L$ nor $D_S$. 
Therefore, $a_1$ must vanish. 
It should be noted that this subtle argument is not true 
of Eq. ($\ref{g-expansion}$), 
since there are a lot of ways for combining terms 
in an infinite series. 

As a result, we find only $g(x) = a_0$, which leads to 
$f(x) = a_0/x$. 
This equation states exactly that 
the modified bending angle is 
proportional to the inverse square of the impact parameter 
(e.g., in Eq. ($\ref{tildealpha}$)). 
One example of such inverse-square corrections is 
the second-order approximation of the Schwarzschild lens 
\cite{EOAK,ES}. 
Another example in general relativity is 
the Reissner-Nordstrom solution representing a charged black hole, 
for which the correction to the bending angle is $\sim q^2/b^2$ 
for a small charge $q$. 

Note that there is a subtle but large difference 
between treatments of $a_1$ and $a_0$. 
This is because the deflection angle depends on 
the lens mass and impact parameter 
but not any other distances such as $D_{LS}$ 
in the lensing theory 
based on the null infinity condition \cite{SEF}. 
The coefficient of $a_1$ is given by Eq. (\ref{a1}) 
as the {\it ill-defined} sum of the (nondimensional) unity and 
the combination of the mass and distances, 
so that the distances cannot be removed 
by using the impact parameter. 
On the other hand, $a_0$ has a coefficient as the inverse 
of the impact parameter squared. 
Hence, the coefficient is well-defined and thus allowed.

\subsection{Discussion}
Do modified magnification factors which are discussed above 
have any implications for observations? 
In microlensing, $u$ has a definite dependence 
on time $t$, which is expressed as $u = (u_0^2 + v_T^2 t^2)^{1/2}$ 
for $u_0$ corresponding to the minimum of the impact parameter 
and $v_T$ denoting the transverse angular velocity 
of the source with respect to the lens. 
The modified total amplification thus has 
a different dependence on time and thus produces 
a different shape of the light curve. 
It should be noted that the peak height of the light curve 
is {\it less} informative on modifications, 
because it depends on many parameters 
and hence there is a degeneracy in parameter spaces. 
Shapes of light curves are of greater importance. 
In principle, accurate measurements of light-curve shapes 
can be used for detecting (or observationally constraining) 
the modifications discussed above. 

\begin{figure}[t]
\includegraphics[width=9.0cm]{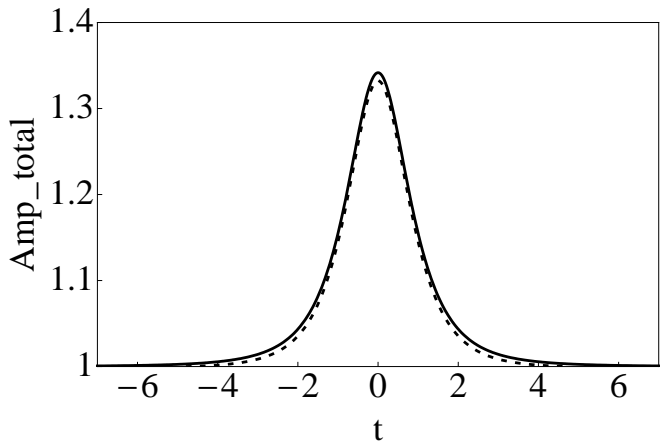}
\caption{
Light curves for $u_0 =1$: 
The solid curve is based on the total amplification 
in the standard form. 
The dashed one includes the Yukawa-type correction with 
$\varepsilon = 0.2$ and $\tilde{K} = 1$. 
Time $t$ is in units of the Einstein cross time 
as $\theta_E / v_T$. 
}
\label{fig1}
\end{figure}

\begin{figure}[t]
\includegraphics[width=9.0cm]{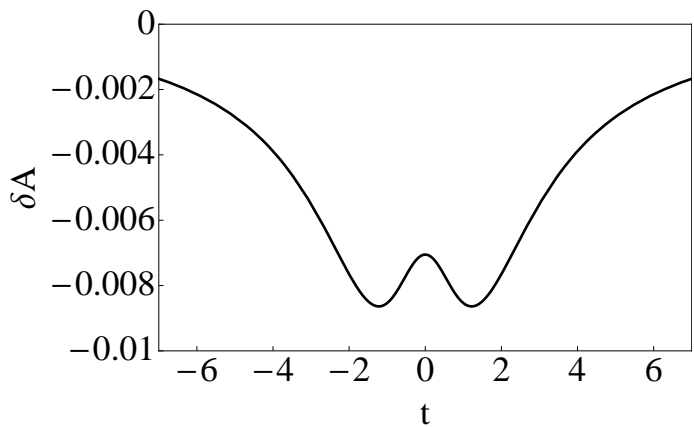}
\caption{
The difference of the two light curves in Figure \ref{fig1}: 
This shows the dependence of $\delta A$ on time 
(through the impact parameter). 
}
\label{fig2}
\end{figure}

For instance, we take a correction to the deflection angle 
as the Yukawa-type function as 
\begin{equation}
\frac{F(b)}{b} = \frac{e^{-K b}}{b} , 
\label{yukawa} 
\end{equation}
where $K$ is roughly the inverse of the characteristic length 
in a modified gravity theory or 
the inverse of a new mass scale. 
Figures $\ref{fig1}$ and $\ref{fig2}$ show effects of the modified 
bending angle on light curves. 
Figure $\ref{fig1}$ gives shapes of light curves for two cases. 
One is based on the amplification factor in the standard form 
and the other is for the Yukawa-type modification 
as $f(\tilde{b}) = \exp(-\tilde{K}\tilde{b})$. 
In order to make it clearer  
the deviation from the standard shape of the light curve, 
we define the ratio as 
\begin{equation}
\delta A(u) \equiv 
\frac{A_{\mbox{total}}^{'} - A_{\mbox{total}}}{A_{\mbox{total}}} , 
\end{equation}
where $A_{\mbox{total}}$ and $A_{\mbox{total}}^{'}$ denote 
the standard form of total amplification  
and the modified one, respectively. 
Figure $\ref{fig2}$ plots $\delta A$, the deviation from 
the standard shape of the light curve. 
For these figures, we choose $u_0 = 1$, $\varepsilon = 0.2$ 
and $\tilde{K} = 1$ so that effects by the modification can be 
distinguished by eye. 

For this model, $\alpha$ is {\it effectively} increased 
and hence $A_{\mbox{total}}$ seems to be enhanced. 
However, $A_{\mbox{total}}$ is decreased as shown by
Fig. $\ref{fig2}$. 
This is because the amplification is caused by 
not the bending angle but its {\it derivative}. 
In fact, the Yukawa-type correction gives 
$df(\tilde{b})/d\tilde{b} 
= - \tilde{K} \exp(-\tilde{K}\tilde{b}) < 0$, 
namely a minus contribution to the amplification factor. 
So far, no systematic deviation in microlensing observations 
has been reported. 
One can thus put a constraint on $K$ as 
$K^{-1} > 1000$ AU $\sim 10^{14}$ m 
for $\varepsilon \sim 1$, 
where we assume 
the photometric accuracy comparable to 0.1 percents 
and $b \sim D_L \theta_E \sim O(1\mbox{AU})$. 
This bound is consistent with gravitational experiments 
in the solar system. 

Finally, we mention an asymptotically massless spacetime, 
for which the above theorem still stands by taking $M \to 0$ 
and considering the linear order in $\varepsilon$. 
One example is Ellis' wormhole that makes a difference 
in light curves \cite{Abe}, since 
the bending angle for this spacetime is $\sim b^{-2}$ at 
the leading order but $\sim b^{-4}$ at the next order.

\section{Conclusion}
We investigated corrections to the conventional lens equation 
in terms of differentiable functions, so that they can express 
not only the second-order effects of $GM$ 
in general relativity but also 
modified gravity theories. 
It was shown that, 
provided that the spacetime is static, spherically symmetric 
and asymptotically flat, 
the total amplification by microlensing 
remains unchanged at the linear order of the correction 
to the deflection angle, 
if and only if the correction takes 
a particular form as the inverse square of the impact parameter, 
whereas the magnification factor for each image is corrected. 
It is concluded that 
the light curve shape by microlensing is inevitably 
changed and will thus allow us to probe modified gravity,  
unless a modification to the deflection angle 
takes the particular form. 

It is left as a future work to use the present formulation 
to probe modified gravity models by microlensing observations. 
Microlensing in our galaxy is sensitive to gravity at short scale 
around a few AU, whereas cosmological microlensing is more useful 
for the large distance physics \cite{SEF}.

\section*{Acknowledgements}
The author would like to thank Fumio Abe and Masumi Kasai 
for fruitful conversations.
This work was supported in part (H.A.) 
by a Japanese Grant-in-Aid 
for Scientific Research from the Ministry of Education, 
No. 21540252.

%

\end{document}